\documentclass[12pt]{article}
\usepackage[a4paper, total={7in, 10in}]{geometry}
\usepackage[parfill]{parskip}
\usepackage{physics, tensor, float, subcaption}
\usepackage{graphicx}
\graphicspath{ {Plots/} }
\usepackage{jhep-mod}
\usepackage{bm}
\usepackage{soul}
\usepackage{amssymb,amsmath,amsthm}
\usepackage{mathrsfs}
\usepackage[utf8]{inputenc}
\usepackage{enumerate}
\usepackage{bigints}
\usepackage{xcolor}
\usepackage{appendix}
\usepackage{graphicx}
\usepackage{float}
\usepackage{tikz}
\usepackage{setspace}
\usepackage{cancel}
\usepackage{doi}
\definecolor{purple}{rgb}{1,0,1}
\definecolor{lime}{HTML}{A6CE39} 

\definecolor{lime}{HTML}{A6CE39}
\newcommand{\orcidicon}{%
	\begin{tikzpicture}
	\draw[lime, fill=lime] (0,0) 
		circle [radius=0.16] 
		node[white] {{\fontfamily{qag}\selectfont \tiny ID}};
	\draw[white, fill=white] (-0.0625,0.095) 
		circle [radius=0.007];
	\end{tikzpicture}
	\hspace{-5mm}
}

\newcommand\orcidMatt{{\href{https://orcid.org/0000-0003-1088-6485}{\orcidicon}}}

\begin{document}


\title{\huge{
Cosmology in Painlev\'e--Gullstrand coordinates
}}


\author{
\Large
Rudeep Gaur 
{\sf  and} Matt Visser\!\orcidMatt\!}
\affiliation{
School of Mathematics and Statistics, Victoria University of Wellington, 
\\
\null\qquad PO Box 600, Wellington 6140, New Zealand.}
\emailAdd{rudeep.gaur@sms.vuw.ac.nz}
\emailAdd{matt.visser@sms.vuw.ac.nz}

\abstract{
\vspace{1em}

Cosmology is most typically analyzed using standard co-moving coordinates, in which the galaxies are (on average, up to presumably small peculiar velocities) ``at rest'', while ``space'' is expanding. But this is merely a specific coordinate choice; and it is important to realise that for certain purposes other, (sometimes \emph{radically}, different) coordinate choices might also prove useful and informative, but without changing the underlying physics. 
Specifically, herein we shall consider the $k=0$ spatially flat FLRW cosmology but in Painlev\'e--Gullstrand coordinates --- these coordinates are very explicitly \emph{not} co-moving:  ``space'' is now no longer expanding, although the distance between galaxies is still certainly increasing. 
Working in these Painlev\'e--Gullstrand coordinates provides an alternate viewpoint on standard cosmology, and the symmetries thereof, and also makes it somewhat easier to handle cosmological horizons.
With a longer view, we hope that investigating these Painlev\'e--Gullstrand coordinates might eventually provide a better framework for understanding large deviations from idealized FLRW spacetimes. We illustrate these issues with a careful look at the Kottler and McVittie spacetimes.
\bigskip

\bigskip
\noindent
{\sc Date:} 18 July 2022; 14 August 2022; \LaTeX-ed \today


\bigskip
\noindent{\sc PhySH:} 
Gravitation
}

\maketitle
\def\tr{{\mathrm{tr}}}
\def\diag{{\mathrm{diag}}}
\def\cof{{\mathrm{cof}}}
\def\pdet{{\mathrm{pdet}}}
\def\d{{\mathrm{d}}}
\def\L{{\mathcal{L}}}
\parindent0pt
\parskip7pt

\clearpage
\section{Introduction}

Coordinate freedom in general relativity is an extremely powerful tool; but a very subtle one that took almost 45 years for most of the general relativity community to fully internalize.\footnote{Unfortunately, not everyone in the wider cosmology and relativity communities has as yet fully understood coordinate freedom in general relativity.}
 A judicious choice of coordinates can often make some aspect of the physics easy and obvious, but may make other aspects of the physics more obscure. On the other hand, no coordinate choice, (no matter how obtuse), can actually change the underlying physics. 
For instance, at a purely theoretical level, locally geodesic and Riemann normal coordinate systems greatly simplify manipulations leading to the Bianchi identities. At a more physical level, locally geodesic and Riemann normal coordinate systems greatly simplify analysis and understanding of the Einstein equivalence principle. 
See any of a vast number of relevant textbooks for more details on these issues~\cite{Peebles:1994,Weinberg:2008, MTW, Wald, Padmanabhan, Hartle, Carroll, Weinberg, Poisson, Hobson, Adler-Bazin-Schiffer, D'Inverno}.

Herein we will explore some unusual coordinate choices in cosmology. While typically in a cosmological setting one uses comoving coordinates, tied to the average Hubble flow, this is by no means a necessary choice. Choosing non-comoving coordinates, (specifically, a cosmological variant of the Painlev\'e--Gullstrand coordinates, will simplify some aspects of the discussion, while (apparently) making other aspects more complicated, but without changing the underlying physics.
For relevant background on Painlev\'e--Gullstrand coordinates see references~\cite{painleve1, painleve2, gullstrand, Poisson-Martel, Faraoni:2020, heuristic,
Visser:unexpected, analog:1997, analog:2005, Nielsen:2005, PGLT1, PGLT2, PGLT3, PGLT4}.

Explicitly choosing a cosmological variant of the Painlev\'e--Gullstrand coordinates will allow us to eliminate the expansion of ``space'', at the price that typical galaxies will now be represented by ``moving'' Eulerian observers --- the distance between galaxies will still be increasing, there will still be a Hubble flow. 
Furthermore, in these Painlev\'e--Gullstrand coordinates the light cones are ``tipped over'', so that ``faster-than-light'' with respect to non-expanding ``space'' is not the same as ``faster-than-light'' with respect to the locally defined light-cones.
This provides an alternative viewpoint on the Hubble expansion, one that some cosmologists might be more comfortable with. We carefully consider the symmetries of FLRW spacetime, the crucial difference between apparent horizons and causal horizons, and as an example of large deviations from FLRW consider several versions of the Kottler and McVittie spacetimes. Our conventions will be those of Misner--Thorne--Wheeler~\cite{MTW}.

\clearpage
\section{Spatially flat FLRW cosmology}

Overwhelming observational evidence points to the spatially flat  $k=0$ FLRW\break cosmology as being an excellent \emph{zeroth-order} approximation to the very large-scale 
structure of spacetime --- beyond the scale of statistical homogeneity~\cite{Peebles:1994,Weinberg:2008,Weinberg, MTW}. Ultimately one might be interested in investigating \emph{large} non-perturbative deviations from FLRW cosmology~\cite{CFLRW, Szekeres:1974, Garcia-Bellido:2008,Marra:2007,Garfinkle:2006, timescape1, timescape2, timescape3}, but for now we shall focus on the idealized case of exact FLRW spacetime. 
Standard presentations of FLRW spacetime can be found in many places, see for instance~\cite{Peebles:1994,Weinberg:2008, MTW, Wald, Padmanabhan, Hartle, Carroll, Weinberg, Hobson, Poisson, Adler-Bazin-Schiffer, D'Inverno}.
Let us start by considering several useful coordinate systems.

\subsection{Standard comoving coordinates}
\subsubsection{Spherical polar version}

The most common presentation of the spatially flat $k=0$ FLRW cosmology is in terms of the explicit line element
\begin{equation}
\label{E:FLRW1}
\d s^2 = - \d t^2 + a(t)^2 \{ \d r^2 + r^2 \d\Omega^2\},
\end{equation}
where $\d\Omega^2=\d \theta^2+\sin^2\theta\;\d\phi^2$. 
In these coordinates the $t$ coordinate is the physical time measured by a fiducial observer of normalized 4-velocity $V^a =(1, 0,0,0)$, so that $V_a = (-1,0,0,0)$, whereas the purely radial ingoing and outgoing light rays are described by the time-dependent opening angle 
\begin{equation}
\left| \d r \over \d t\right| = {1\over a(t)}.
\end{equation}
In these coordinates the $t= (constant)$ spatial slices are 3-flat but \emph{expanding}
\begin{equation}
\label{E:FLRW-3-space-1}
\d s_3^2 =   a(t)^2\{\d  r^2 +  r^2 \d\Omega^2\}.
\end{equation}
For the spatially flat case $k=0$ one has a \emph{choice} as to whether the coordinate $r$ is dimensionless while the scale factor $a$ has units of length, or \emph{vice versa}. For nonzero spatial curvature, if one sets  $k=\pm 1$ then one is forced to take the coordinate $r$ to be dimensionless, while the scale factor $a$ has units of length. We shall make the same choice in the spatially flat $k=0$ case. 

\clearpage
\subsubsection{Cartesian version}
We could equally well use comoving Cartesian coordinates for the spatial slices
\begin{equation}
\label{E:FLRW2}
\d s^2 = - \d t^2 + a(t)^2 \{ \d x^2 + \d y^2 + \d z^2\}.
\end{equation}
Thence
\begin{equation}
\label{E:FLRW-3-space-2}
\d s_3^2 =   a(t)^2\{\d x^2 + \d y^2 + \d z^2\}.
\end{equation}
Doing so will not change the physics, just the presentation.\footnote{There are already examples, thankfully extremely rare, of people who cannot successfully handle the spherical-polar to Cartesian conversion.}
For instance the light cones are now described by the time-dependent opening angle 
\begin{equation}
\left| \d \vec x \over \d t\right| = {1\over a(t)}.
\end{equation}
For the fiducial observers we still have the normalized 4-velocity $V^a =(1, 0,0,0)$, so that $V_a = (-1,0,0,0)$.

Let us now consider some other possible coordinate systems.

\subsection{Conformal time coordinate}

Define a conformal time coordinate by
\begin{equation}
\eta(t)  = \int_0^t {\d \bar t \over a(\bar t)}.
\end{equation}
Note that with our conventions the scale factor $a$ has units of distance so that the conformal time is dimensionless. 
One can formally invert this definition to obtain $t(\eta)$, and thereby implicitly define $a(\eta) = a (t(\eta))$.

\subsubsection{Spherical polar version}

Using conformal time we can  re-cast the line element as
\begin{equation}
\label{E:FLRW3}
\d s^2 =  a(\eta)^2 \{ -\d \eta^2 + \d r^2 + r^2 \d\Omega^2\}.
\end{equation}
This choice of coordinate system makes manifest the fact that $k=0$ FLRW spacetime is conformally flat (the Weyl tensor is identically zero).\footnote{One can still to this day encounter a non-empty set of people who insist that the conformal metric is physically different from the standard one.}

This conformal time coordinate has the technical advantage that the radial ingoing and outgoing light rays are now particularly simple
\begin{equation}
\left| \d r \over \d \eta\right| = {1}.
\end{equation}
In contrast, the proper time (clock time) measured by a fiducial observer, now with normalized 4-velocity $V^a = {1\over a(\eta)}\;(1,0,0,0)$, becomes more complicated. Note that for the related co-vector one now has $V_a= a(\eta) \; (-1,0,0,0)$.
For the proper time one has
\begin{equation}
\tau(\eta) = \int_0^\eta a(\bar\eta) \; \d\bar\eta.
\end{equation}
This is a common theme of coordinate freedom --- coordinates can often be chosen to make \emph{some} formulae simpler, (in this case, the light cones),  at the cost of complicating \emph{other} formulae (in this case, the proper time). 

\subsubsection{Cartesian version}

We could equally well use comoving Cartesian coordinates for the spatial slices and re-write (\ref{E:FLRW3}) as
\begin{equation}
\label{E:FLRW4}
\d s^2 = a(\eta)^2 \{ - \d \eta^2 +  \d x^2 + \d y^2 + \d z^2\}.
\end{equation}
Adopting these coordinates will not change the physics.
The light cones are now particularly simple
\begin{equation}
\left| \d \vec x \over \d \eta\right| = {1}.
\end{equation}
This simplified light cone structure makes the causal structure in these conformal coordinates particularly easy to deal with.
For the fiducial observers we again have both $V^a = {1\over a(\eta)}\;(1,0,0,0)$ and $V_a= a(\eta) \; (-1,0,0,0)$.

\subsection{Painlev\'e--Gullstrand coordinates}

We shall now introduce the cosmological Painlev\'e--Gullstrand coordinate systems that are of central importance to this article. (For relevant background discussion see references~\cite{painleve1, painleve2, gullstrand, Poisson-Martel, Faraoni:2020, heuristic, analog:1997,  Visser:unexpected, analog:2005, Nielsen:2005, PGLT1, PGLT2, PGLT3, PGLT4}.)

\subsubsection{Spherical polar version}

\enlargethispage{20pt}
\paragraph{Metric:}
Starting from the standard line element (\ref{E:FLRW1}), let us now make the time-dependent coordinate transformation $\bar r = a(t) \, r$. Then $\bar r$ is a Schwarzschild radial coordinate, based on the notion of area, since the area of a 2-sphere of coordinate radius $\bar r$ is simply $4\pi \, \bar r^2$. (Consequently, these are sometimes called ``area coordinates''.) 

\clearpage
Furthermore 
\begin{equation}
\d\bar r = a(t) \,\d r + r \,\dot a(t) \,\d t = a(t)\, \d  r + H(t) \, \bar r \,\d t,
\end{equation}
where $H(t) =\dot a(t)/a(t)$ is the Hubble parameter. 
Thence 
\begin{equation}
a(t)\, \d  r= \d\bar r - H(t) \, \bar r \,\d t.
\end{equation}
Consequently in these coordinates the line element becomes
\begin{equation}
\label{E:FLRW5}
\d s^2 = - \d t^2 + \{ [\d \bar r - H(t)\, \bar r \,\d t]^2 + \bar r^2 \d\Omega^2\}.
\end{equation}

That is
\begin{equation}
\label{E:FLRW5b}
\d s^2 = - (1-H(t)^2 \,\bar r^2)\, \d t^2 -2 H(t) \,\bar r \,\d\bar r\,\d t+ \{ \d \bar r^2 + \bar r^2 \d\Omega^2\}.
\end{equation}
Note that the line element only contains the scale factor \emph{implicitly}, via the Hubble parameter $H(t)$. 
Furthermore, in these new coordinates the $t= (constant)$ spatial slices are again 3-flat, but are now \emph{non-expanding}
\begin{equation}
\label{E:FLRW-3-space-5}
\d s_3^2 =   \d \bar r^2 + \bar r^2 \d\Omega^2.
\end{equation}
Adopting ADM terminology, (see for instance references~\cite{ADM1,ADM2}), all the non-trivial aspects of the $k=0$ FLRW spacetime geometry have now been pushed into the shift vector $g_{0i} = (-H(t) \,\bar r, 0, 0)$. The lapse function is still unity, one still has  $N^2=-g^{tt} = 1$.
Coordinate systems of this type are called Painlev\'e--Gullstrand coordinates~\cite{painleve1, painleve2, gullstrand, Poisson-Martel, Faraoni:2020, heuristic}. 
Very many, (but certainly not all), physically interesting spacetimes can be put into this Painlev\'e--Gullstrand form. For example: all of the Schwarzschild spacetime~\cite{analog:1997, analog:2005}, most of the Reissner--Nordstr\"om spacetime (the region $r>{Q^2\over2m}$), all of the Lense--Thirring spacetime~\cite{PGLT1,PGLT2,PGLT3,PGLT4}, all spherically symmetric spacetimes (at least locally) can be recast in this form;\footnote{In spherical symmetry the only obstructions to the global existence of Painlev\'e--Gullstrand coordinates are the possible existence of wormhole throats, (since then the area radial coordinate cannot be monotone), and/or negative Misner--Sharp quasi-local mass~\cite{Faraoni:2020}, (since then the shift vector is forced to become imaginary). } but not the Kerr or Kerr--Newman spacetimes~\cite{Valiente-Kroon:2003,Valiente-Kroon:2004}.  \enlargethispage{20pt}

In these Painlev\'e--Gullstrand coordinates there is manifestly an \emph{apparent} horizon, (where $g_{tt}=0$), at the Hubble radius $\bar r_\mathrm{Hubble} = 1/H(t)$. Additionally, the fiducial Eulerian (geodesic) observers have covariant 4-velocity $V_a=(-1,0,0,0)$, which now corresponds to the contravariant 4-velocity $V^a = (1,\, H(t) \, \bar r, 0,0)$.  So a typical galaxy (ignoring peculiar velocities) is certainly ``moving'' in this coordinate system. While ``space'' is now non-expanding, the  Hubble flow is explicit, with $V^r = H(t) \, \bar r$.
We emphasize, yet again, that this is just a coordinate choice --- no physics has been harmed in making this choice of coordinates. 

\clearpage
The radial ingoing and outgoing light rays are now described  by
\begin{equation}
\left|{\d \bar r\over \d t} - H(t)\, \bar r \right| = 1.
\end{equation} 
That is
\begin{equation}
{\d \bar r\over \d t} =  H(t)\, \bar r  \pm  1,
\end{equation} 
whereas a typical galaxy (vanishing peculiar velocity) is moving with 3-velocity
\begin{equation}
{\d \bar r\over \d t} =  H(t)\, \bar r,
\end{equation} 
which safely lies inside the light cone. 

\paragraph{Tetrad:}
\enlargethispage{30pt}
A suitable co-tetrad is easily read off from the line element:
\begin{equation}
e^{\hat t}{}_a = (1,0,0,0); \quad 
e^{\hat r}{}_a = (-H\bar r, 1,0,0); \quad
e^{\hat \theta}{}_a = (0,0,\bar r,0); \quad
e^{\hat \phi}{}_a = (0,0,0,\bar r \sin\theta).
\end{equation}
The corresponding tetrad is then given by the timelike leg
\begin{equation}
e_{\hat t}{}^a = V^a =(1, H(t) \, \bar r, 0,0); \quad 
\end{equation}
and the particularly simple spatial triad
\begin{equation} 
e_{\hat r}{}^a = (0, 1,0,0); \quad
e_{\hat \theta}{}^a = \left(0,0,{1\over \bar r},0\right); \quad
e_{\hat \phi}{}^a = \left(0,0,0,{1\over\bar r \sin\theta}\right).
\end{equation}
It is easy to check that with $\eta_{\hat m\hat n} = \mathrm{diag}\{-1,1,1,1\}$ one has (as expected):
\begin{equation}
g_{ab} = \eta_{\hat m\hat n}\;\; e^{\hat m}{}_a \;\; e^{\hat n}{}_b;
\qquad
\eta_{\hat m\hat n} =  g_{ab} \;\; e_{\hat m}{}^a \;\; e_{\hat n}{}^b.
\end{equation}
A brief computation yields the orthonormal components of the Riemann tensor
\begin{equation}
R_{\hat t\hat r\hat t\hat r} = 
R_{\hat t\hat \theta\hat t\hat \theta} =
R_{\hat t\hat \phi\hat t\hat \phi} =
 - H^2-\dot H = - {\ddot a\over a}; 
\qquad
R_{\hat r\hat \theta\hat r\hat \theta} = 
R_{\hat r\hat \phi\hat r\hat \phi} =
R_{\hat \theta\hat \phi\hat \theta\hat \phi} =
H^2. 
\end{equation}
The Weyl tensor is (as expected) identically zero, while for the Einstein and Ricci tensors one has
\begin{equation}
G_{\hat t\hat t} =  3 H^2; \qquad 
G_{\hat r\hat r} = G_{\hat\theta\hat\theta} = G_{\hat\phi\hat\phi} 
=  - 3 H^2-2 \dot H;
\end{equation}
and
\begin{equation}
R_{\hat t\hat t} =  -3 H^2 -3\dot H= -3\, {\ddot a\over a}; \qquad 
R_{\hat r\hat r} = R_{\hat\theta\hat\theta} = R_{\hat\phi\hat\phi} 
=   3 H^2 +\dot H.
\end{equation}

The Ricci scalar is $R = 12 H^2+6 \dot H$. 

\clearpage
These orthonormal components (that is, components in the basis defined by the orthonormal tetrad) are identical (as they should be) to the standard orthonormal components defined in comoving coordinates. 
The cosmological Friedmann equations will be unaffected. 
The fiducial observers, with 4-velocity $V^a =(1, H(t) \, \bar r, 0,0)$ are geodesic.

\subsubsection{Cartesian version}
One could also construct a Cartesian version of Painlev\'e--Gullstrand coordinates.

\paragraph{Metric:}
Define
\begin{equation}
\bar x= \bar r \,\sin\theta\,\cos\phi; \qquad
\bar y= \bar r \,\sin\theta\,\sin\phi; \qquad
\bar z= \bar r \,\cos\theta.
\end{equation}
Then $\bar r=\sqrt{\bar x^2 +\bar y^2+\bar z^2}$, and our spherical polar Painlev\'e--Gullstrand version of $k=0$ FLRW spacetime,
\begin{equation}
\label{E:FLRW5bb}
\d s^2 = - (1-H(t)^2 \,\bar r^2) \d t^2 -2 H(t) \,\bar r\,\d\bar r \,\d t+ \{ \d \bar r^2 + \bar r^2 \d\Omega^2\},
\end{equation}
now becomes
\begin{equation}
\label{E:FLRW6}
\d s^2 = - (1-H(t)^2 \,\{\bar x^2 +\bar y^2+\bar z^2\}) \d t^2 
-2 H(t) \,\{\bar x\,\d\bar x + \bar y\,\d\bar y +\bar z\,\d\bar z \} \,\d t+ \{ \d \bar x^2 + \d \bar y^2 + \d\bar z^2\}.
\end{equation}
That is
\begin{equation}
\label{E:FLRW7}
\d s^2 = -  \d t^2 
 + \left\{ [\d \bar x- H(t) \,\bar x\, \d t]^2 
 + [\d \bar y - H(t) \,\bar y \,\d t]^2 
 + [\d\bar z - H(t) \,\bar z \, \d t]^2\right\}.
\end{equation}
In 3-vector notation the line element is 
\begin{equation}
\label{E:FLRW7b}
\d s^2 = -  \d t^2 
 + \left[\d \vec{\bar x}- H(t) \,\vec{\bar x}\, \d t\right]^2.
\end{equation}
The light cones are now simply
\begin{equation}
\left| {\d \vec{\bar x}\over \d t} - H(t) \,\vec{\bar x}\right| = 1.
\end{equation}
That is
\begin{equation}
 {\d \vec{\bar x}\over \d t} = H(t) \,\vec{\bar x} + \hat n,
\end{equation}
where $\hat n$ is an arbitrary unit vector in 3-space. 
\enlargethispage{30pt}

Since a typical galaxy (zero peculiar velocity) is moving with 3-velocity
\begin{equation}
 {\d \vec{\bar x}\over \d t} = H(t) \,\vec{\bar x},
\end{equation}
the Hubble flow lies safely inside the light cones.

\clearpage
\paragraph{Tetrad:}
A suitable co-tetrad is easily read off from the line element:
\begin{equation}
e^{\hat t}{}_a = (1,0,0,0); \quad 
e^{\hat x}{}_a = (-H\bar x, 1,0,0); \quad
e^{\hat y}{}_a = (-H\bar y,0,1,0); \quad
e^{\hat z}{}_a = (-H\bar z,0,0,1).
\end{equation}
The corresponding tetrad is thus:
\begin{equation}
e_{\hat t}{}^a = (1, H\bar x,H\bar y,H\bar z); \quad 
e_{\hat x}{}^a = (0, 1,0,0); \quad
e_{\hat y}{}^a = \left(0,0,1,0\right); \quad
e_{\hat z}{}^a = \left(0,0,0,1\right).
\end{equation}
Note how simple the spatial triad now is: $e_{\hat i}{}^j = \delta_i{}^j$. 
A brief computation yields the orthonormal components of the Riemann tensor
\begin{equation}
R_{\hat t\hat i\hat t\hat j} = 
-\{H^2+\dot H\}\, \delta_{ij} = -{\ddot a\over a}\, \delta_{ij}; 
\qquad
R_{\hat i\hat j\hat k\hat l} = 
H^2 \{ \delta_{ik} \delta_{jl}- \delta_{il} \delta_{jk}\}. 
\end{equation}

The Weyl tensor is (as expected) still identically zero, while for the Einstein and Ricci tensors one has
\begin{equation}
G_{\hat t\hat t} =  3 H^2; \qquad 
G_{\hat i\hat j} =  -\{3 H^2+ 2 \dot H \} \delta_{ij};
\end{equation}
and
\begin{equation}
R_{\hat t\hat t} =  -3\dot H -3 H^2= -3 \, {\ddot a\over a}; \qquad 
R_{\hat i\hat j} = \{\dot H + 3 H^2\} \delta_{ij}.
\end{equation}

The Ricci scalar is still $R = 6 \dot H +12 H^2$. 

These orthonormal components (that is, components in the basis defined by the orthonormal tetrad) are identical (as they should be) to the standard orthonormal components defined in the usual comoving coordinates. Consequently the cosmological Friedmann equations will be unaffected. 
The fiducial observers, with 4-velocity 
$V^a =(1,\, H(t) \, \bar x, H(t) \, \bar y,H(t) \, \bar z)$ are geodesic.

\subsection{Summary}

Cosmological  Painlev\'e--Gullstrand coordinates, (appropriate to $k=0$ FLRW spacetime), have some very nice features. Three-space is flat and non-expanding --- but the price one pays for this is that the light cones are ``tipped over'' and that the galaxies are ``moving'' with respect to ``space''. 

The Hubble flow is then very explicit
\begin{equation}
 {\d \vec{\bar x}\over \d t} = H(t) \,\vec{\bar x}.
\end{equation}
The light cones are characterized by 
\begin{equation}
 {\d \vec{\bar x}\over \d t} = H(t) \,\vec{\bar x} + \hat n,
 \qquad 
 |\hat n|=1.
\end{equation}
There is as always a ``conservation of difficulty'' inherent in any coordinate choice; since the underling physics cannot change.

\section[Symmetries of spatially flat FLRW:
Explicit, partial, and hidden]
{Symmetries of spatially flat FLRW: \\
\null\qquad Explicit, partial, and hidden}

The FLRW spacetime possesses a number of explicit symmetries (associated with Killing vectors) and hidden symmetries (associated with Killing tensors and Killing--Yano 2-forms). These symmetries are often more obvious in appropriately chosen coordinates. We present several examples below. 

\subsection{Killing vectors}

The explicit symmetries of FLRW spacetime are associated with the rotational and translational Killing vectors.

\subsubsection{Spherical symmetry}

The 2-sphere  $S^2$, with metric $\d s^2 = \d\theta^2 + \sin^2\theta\, \d\phi^2$, can be shown to have  an over-complete set of linearly dependent (rotational) Killing vectors.
They can most easily be chosen to be (see for instance~\cite[page 139]{Carroll}):
\begin{equation}
R_1= -\sin\phi \;\partial_\theta - {\cos\phi\over\tan\theta}\; \partial_\phi; \qquad
R_2= \cos\phi \;\partial_\theta - {\sin\phi\over\tan\theta} \;\partial_\phi; \qquad
R_3= \partial_\phi;
\end{equation}
and are subject to the constraint
\begin{equation}
(\cos\phi\,\tan\theta) R_1 + (\sin\phi\,\tan\theta) R_2+  R_3 =0.
\end{equation}
It is easy to check that these three vectors all satisfy Killing's equation, that is  $[R_{\{1,2,3\}}]_{(a;b)}=0$. 

\clearpage
Note that $R_3$ is particularly simple; and has the obvious physical  interpretation of corresponding to a translation in the azimuthal $\phi$ coordinate; a rotation around the poles located at $\theta\in\{0,\pi\}$. In counterpoint $R_1$ and $R_2$ at first look a little more complicated,  but there is no substantial difference; they correspond to rotations around the points $(\theta=\pi/2;\phi\in\{0,\pi\})$ and $(\theta=\pi/2;\phi\in\{\pi/2,3\pi/2\})$ respectively. 
(These are the points where the Killing vectors $R_1$ and $R_2$ vanish.)
These Killing vectors defined on $S^2$ can then be bootstrapped without alteration into the generic spherically symmetric 3-space: 
$\d s^2 = g_{rr}(r) \, \d r^2 + r^2(\d\theta^2 + \sin^2\theta\, \d\phi^2)$.

Specifically, flat 3-space in Cartesian coordinates, with line element  given by $\d s^2 = \d x^2+\d y^2+\d z^2$, is also spherically symmetric and also exhibits an over-complete set of linearly dependent (rotational) Killing vectors:
\begin{equation}
R_1= y \,\partial_z - z\,\partial_y; \qquad
R_2= z \,\partial_x - x\,\partial_z; \qquad
R_3= x \,\partial_y - y\,\partial_x;
\end{equation}
subject to the constraint
\begin{equation}
x \,R_1 + y \,R_2+ z \,R_3 =0.
\end{equation}
This presentation makes manifest the intimate relationship between the (rotational) Killing vectors and the angular momentum operators of quantum mechanics. 
(For some specific purposes we see that Cartesian coordinates are clearly superior to spherical polar coordinates.) 
These Killing vectors can then be bootstrapped into the (3+1) dimensional FLRW spacetime; in any of the various coordinate systems discussed above.

\subsubsection{Spatial translation symmetry}

The FLRW spacetimes also possess 3 linearly independent spatial translation Killing vectors. For the $k=0$ FLRW spacetime in standard comoving Cartesian coordinates, where  one has $\d s^2 = - \d t^2 +  a(t)^2 \{ \d x^2 +\d y^2 +\d x^2\}$, these spatial translation Killing vectors are simply
\begin{equation}
T_1 = \partial_x; \qquad T_2=\partial_y; \qquad T_3 = \partial_z.
\end{equation}
But, since $\bar x^i = a(t) x^i$, and we want to find the translation Killing vectors for the Painlev\'e--Gullstrand form of FLRW
\begin{equation}
\label{E:FLRW7b}
\d s^2 = -  \d t^2 
 + \left\{ [\d \bar x- H(t) \,\bar x\, \d t]^2 
 + [\d \bar y - H(t) \,\bar y \,\d t]^2 
 + [\d\bar z - H(t) \,\bar z \, \d t]^2\right\},
\end{equation}
we observe that
\begin{equation}
{\partial\over\partial x^i} = 
{\partial \bar x^a\over\partial x^i} 
 {\partial\over\partial \bar x^a}
= 
{\partial \bar x^j\over\partial x^i}  {\partial\over\partial \bar x^j}
+ \left.{\partial t\over\partial x^i}\right|_{\bar x}   {\partial\over\partial t}
= a(t) {\partial\over\partial \bar x^i}\,.
\end{equation}
So in the Painlev\'e--Gullstrand  Cartesian coordinate system the space translation Killing vectors are
\begin{equation}
T_1 = a(t) \,\partial_{\bar x}; \qquad 
T_2= a(t) \,\partial_{\bar y}; \qquad 
T_3 = a(t) \,\partial_{\bar z}.
\end{equation}

If one wishes instead to use comoving spherical polar coordinates then the spatial translation Killing vectors appear to be somewhat less intuitive
\begin{eqnarray}
T_1 = \partial_x &=&  
{\partial x^a\over \partial x}\, {\partial\over \partial x^a} =
\sin\theta\,\cos\phi\; \partial_r
+{\cos\theta\,\cos\phi\over r} \;\partial_\theta
- {\sin\phi \over r\sin\theta} \;\partial_\phi\,;
\\
T_2=\partial_y &=& 
{\partial x^a\over \partial y}\, {\partial\over \partial x^a} =
\sin\theta\,\sin\phi \;\partial_r
+{\cos\theta\,\sin\phi\over r} \;\partial_\theta
 +{\cos\phi \over r\sin\theta} \;\partial_\phi\,;
\\
T_3 = \partial_z &=& 
{\partial x^a\over \partial z}\, {\partial\over \partial x^a} =
\cos\theta \;\partial_r
-{\sin\theta\over r} \; \partial_\theta\,.
\end{eqnarray}
That is: While one can certainly use spherical polar coordinates to describe the spatial translations, after all, it's just a coordinate change, it is perhaps unsurprising that the relevant  Killing vectors then (superficially) appear to be somewhat more complicated.

Similarly if one wishes  to use Painlev\'e--Gullstrand spherical polar coordinates then the spatial translation Killing vectors are 
\begin{eqnarray}
T_1 = a(t) \, \partial_{\bar x} &=&  a(t) \,\left\{ 
\sin\theta\,\cos\phi\; \partial_{\bar r}
+{\cos\theta\,\cos\phi\over \bar r} \;\partial_\theta
- {\sin\phi \over \bar r\sin\theta} \;\partial_\phi\right\};
\\
T_2= a(t) \,\partial_{\bar y} &=& a(t) \,\left\{
\sin\theta\,\sin\phi \;\partial_{\bar r}
+{\cos\theta\,\sin\phi\over \bar r} \; \partial_\theta
 +{\cos\phi \over \bar r\sin\theta} \; \partial_\phi\right\};
\\
T_3 =  a(t) \,\partial_{\bar z} &=& a(t) \,\left\{
\cos\theta \;\partial_{\bar r}
-{\sin\theta\over \bar r} \; \partial_\theta\right\}.
\end{eqnarray}
In short, for some purposes the use of spherical polar coordinates is less useful than one might hope.

\subsubsection{Time translation not-quite symmetry}

Since the FLRW spacetime is explicitly time dependent there is no Killing vector for time translations --- however one does have the next best thing --- a conformal Killing vector for time translations. Specifically the timelike co-vector $T^\flat = -a(t) \, \d t$, that is $T_a = -(a(t),0,0,0)$, which in comoving coordinates has vector components $T^a = a(t)\; (1,0,0,0)$, and in Painlev\'e--Gullstrand  coordinates has  vector components $T^a = a(t) (1,\,H\bar x,H\bar y,H\bar z)$, is a conformal Killing vector which satisfies
\begin{equation}
\L_{T}\, g = \dot a(t) \; g\,.
\end{equation}

\clearpage
Explicitly
\begin{equation}
T_{(a;b)} = \dot a(t) \; g_{ab}\,.
\end{equation}
This is enough to guarantee a conservation law for affinely parameterized null geodesics 
\begin{equation}
a(t)\; {\d t \over \d\lambda} = \hbox{(constant)}.
\end{equation}
The existence of this timelike conformal Killing vector is ultimately the reason why the locally measured energy of freely propagating photons is proportional to the inverse of the scale factor
\begin{equation}
E(t)\; a(t) =  \hbox{(constant)}.
\end{equation}
Equivalently, this timelike conformal Killing vector guarantees that the locally measured wavelength of freely propagating photons is proportional to the scale factor
\begin{equation}
\lambda(t) \propto a(t).
\end{equation}
The existence of this timelike conformal Killing vector in FLRW spacetimes is often not emphasized or explained in pedagogical presentations, but is central to understanding photon propagation over cosmological distances.

\subsection{Killing tensors}

In addition to the obvious Killing vectors (corresponding to rotations and spatial translations), and the trivial Killing tensors that one can build out of the metric and the Killing vectors, 
the $k=0$ FLRW geometry possesses two (non-trivial) Killing tensors, which satisfy the 3-index version of Killing's equation $K_{(ab;c)}=0$. 

\subsubsection{Spherical symmetry}

Due to spherical symmetry there is a  non-trivial Killing tensor (see for instance~\cite{PGLT2}) which in comoving spherical polar coordinates takes the form
\begin{equation}
(K_\Omega)_{ab} \;\d x^a \otimes \d x^b = a(t)^4 r^4 \{ \d \theta^2+\sin^2\theta\;\d\phi^2\}
= (a(t)^2 r^2 \;\d\theta)^2 + (a(t)^2 r^2 \sin\theta \; \d \phi)^2.
\end{equation}
In components
\begin{equation}
(K_\Omega)_{ab} = a(t)^2 r^2 \{ g_{ab} + \nabla_a t \, \nabla_b t
- a(t)^2 \,\nabla_a r \,\nabla_b r\}.
\end{equation}
Using the Painlev\'e--Gullstrand  $\bar r$ coordinate, where $\bar r= a(t) r$, one simply has
\begin{equation}
(K_\Omega)_{ab} = \bar r^2 \{ g_{ab} + \nabla_a t \, \nabla_b t
-  (\nabla_a \bar r -H(t) \,\bar r\, \nabla_a t) \,
(\nabla_b \bar r-H(t) \,\bar r \,\nabla_b t)\}.
\end{equation}
Thence in Painlev\'e--Gullstrand  spherical polar coordinates a brief calculation yields 
\begin{equation}
(K_\Omega)_{ab} \;\d x^a \otimes \d x^b = \bar r^4 \{ \d \theta^2+\sin^2\theta\;\d\phi^2\}
= (\bar r^2 \;\d\theta)^2 + (\bar r^2 \sin\theta \; \d \phi)^2 .
\end{equation}

Furthermore in Painlev\'e--Gullstrand  Cartesian coordinates one can write
\begin{equation}
(K_\Omega)_{ab} = \left[\begin{array}{cccc}
0&0&0&0\\
0&\bar y^2+\bar z^2&-\bar x \bar y&-\bar x\bar z \\
0&-\bar x\bar y&\bar x^2+ \bar z^2 &-\bar y\bar z \\
0&-\bar x\bar z&-\bar y\bar z&\bar x^2+\bar y^2 \\
\end{array}\right],
\end{equation}
that is
\begin{equation}
(K_\Omega)_{ab} \;\d x^a \otimes \d x^b = 
(\bar x^2 +\bar y^2+\bar z^2) (\d \bar x^2+\d \bar y^2+\d \bar z^2) 
- (\bar x \,\d \bar x+ \bar y \,\d \bar y +\bar z\,\d\bar z)^2.
\end{equation}
All four of these different coordinate representations of the angular Killing tensor $K_\Omega$ carry the same mathematical and physical information. 
It is easy to check that in any of these situations $K_\Omega$ satisfies the 3-index version of Killing's equation $[K_\Omega]_{(ab;c)}=0$. 

\subsubsection{Spatial translation symmetry}

There is also a non-trivial Killing tensor associated with uniformity of the spatial slices $\Sigma$. See for instance~\cite[page 344]{Carroll}. In comoving coordinates this takes the simple form
\begin{equation}
(K_\Sigma)_{ab} \;\d x^a \otimes \d x^b 
= a(t)^4 \{\d r^2 + r^2(\d\theta^2 + \sin^2\theta\,\d\phi^2\}
=  a(t)^4 \{\d x^2+\d y^2 +\d z^2\}.
\end{equation}
This can also be written as
\begin{equation}
(K_\Sigma)_{ab} = a(t)^2 ( g_{ab} + \nabla_a t \,\nabla_b t).
\end{equation}
Equivalently
\begin{equation}
(K_\Sigma)_{ab} \;\d x^a \otimes \d x^b =
a(t)^2 \left\{ \d s^2 + \d t^2 \right\}.
\end{equation}
Phrased in this way it is clear what happens in 
Painlev\'e--Gullstrand  coordinates.  First, in Painlev\'e--Gullstrand  spherical polar coordinates, from equation (\ref{E:FLRW5}) one has 
\begin{equation}
(K_\Sigma)_{ab} \;\d x^a \otimes \d x^b =
a(t)^2 \left\{ (\d \bar r - H(t) \bar r \d t)^2 + \bar r^2 (\d\theta^2 + \sin^2\theta\,\d\phi^2)  \right\}.
\end{equation}
In contrast, in Painlev\'e--Gullstrand  Cartesian coordinates,
from equation (\ref{E:FLRW7}) one has
\begin{equation}
(K_\Sigma)_{ab} \;\d x^a \otimes \d x^b =
a(t)^2 \left\{ [\d \bar x- H(t) \,\bar x\, \d t]^2 
 + [\d \bar y - H(t) \,\bar y \,\d t]^2 
 + [\d\bar z - H(t) \,\bar z \, \d t]^2 \right\}.
\end{equation}
It is easy to check that in any of these situations $K_\Sigma$ satisfies the 3-index version of Killing's equation $[K_\Omega]_{(ab;c)}=0$.

\subsection{Killing--Yano tensor}

A  Killing--Yano 2-form $Y_{ab} \;\d x^a \wedge \d x^b$ satisfies the differential equation $Y_{a(b;c)}=0$. 
Thence, if we define $K_{ab}= Y_{ae} \,g^{ef}\, Y_{fb}$,
then (with indices between vertical bars not being included in the symmetrization process) we have
\begin{equation}
K_{(ab;c)} =  
Y_{(a|e|;c} \,g^{ef}\, Y_{|f|b)} 
+  Y_{(a|e} \,g^{ef}\, Y_{f|b;c)} = 0+0=0.
\end{equation}
That is, the existence of a Killing--Yano 2-form implies the existence of a 2-index Killing tensor. 

Specifically, the existence of the 
2-index Killing tensor $(K_\Omega)_{ab} \;\d x^a \otimes \d x^b $ that is associated with spherical symmetry is related to the existence of a Killing--Yano 2-form $(Y_\Omega)_{ab} \;\d x^a \wedge \d x^b$. 
In comoving spherical polar coordinates
\begin{equation}
(Y_\Omega)_{ab} \;\d x^a \wedge \d x^b = a(t)^3 r^3 \sin\theta \; \d\theta\wedge\d\phi=
{ (a(t)^2 r^2 \;\d\theta)\wedge(a(t)^2 r^2 \sin\theta \; \d \phi)\over
a(t) r}.
\end{equation}
 In our Painlev\'e--Gullstrand  spherical polar coordinates one simply has
\begin{equation}
(Y_\Omega)_{ab} \;\d x^a \wedge \d x^b = \bar r^3 \sin\theta \; \d\theta\wedge\d\phi=
{ (\bar r^2 \;\d\theta)\wedge(\bar r^2 \sin\theta \; \d \phi)\over
\bar r}.
\end{equation}
The Killing--Yano tensor  is colloquially referred to as the square root of the Killing tensor: $K_{ad}= Y_{ab} \,g^{bc}\, Y_{cd}$. 
Because the Killing--Yano tensor is a 2-form, represented by an anti-symmetric matrix, if it is nonzero it can only have rank 2 or rank 4; which then forces the associated Killing tensor to either have rank 2 or rank 4. Since the 
Killing tensor associated with uniformity of the the spatial slices is manifestly rank 3, that particular Killing tensor will not have an associated Killing--Yano 2-form. 

\subsection{Summary}

FLRW spacetimes possess significant symmetry structure. The spatial and rotational Killing vectors are the most obvious symmetries, but they are far from the only symmetries. The timelike conformal Killing vector can be viewed as an approximate symmetry, one that still leads to a conservation law for null geodesics. 

More subtle are the ``hidden'' symmetries encoded in the non-trivial Killing tensors and the Killing--Yano tensor. 
The specific choice of coordinate system can make some of these symmetries manifest, at the cost of making other symmetries less obvious. 

\section{Cosmological horizons}
Cosmological horizons can be quite tricky to properly define and interpret~\cite{lost}. While event horizons are mathematically clean concepts, and their use underlies many of the singularity theorems, there is a precise technical sense in which any physical observer (represented by a finite-size finite-duration laboratory) cannot ever, \emph{even in principle}, detect an event horizon~\cite{observability}. 
The point is that event horizons are \emph{teleological}, and defining them requires one to back-track from the trump of doom.  
(Quasi-local horizons are much better behaved in this regard; quasi-local horizons can be detected using finite-size finite-duration laboratories.) In the words of Stephen Hawking~\cite{weather} (applied in the context of  black hole physics): 
\begin{quote}``The absence of event horizons means that there are no black holes --- in the sense of regimes from which light can't escape to infinity. There are however apparent horizons which persist for a period of time." 
\end{quote}
Similar, related but distinct,  issues arise in cosmology. One must be very careful to distinguish quasi-local horizons from causal horizons.

\subsection{Apparent horizon (Hubble sphere)}

Consider the 2-sphere located at $\bar r(t)$, with area 
\begin{equation}
S(t) = 4\pi \,\bar r(t)^2,
\end{equation}
and ask how this area evolves as the 2-sphere expands or contracts at the speed of light
\begin{equation}
{\d \bar r\over \d t} =  H(t)\, \bar r  \pm  1.
\end{equation} 
Then for outgoing light rays 
\begin{equation}
\dot S_+(t)  = 8\pi \bar r \left(\d \bar r\over \d t\right)_+  = 
8\pi \bar r \left(  H(t)\, \bar r  +  1 \right) > 0,
\end{equation}
while for ingoing light rays
\begin{equation}
\dot S_-(t)  = 8\pi \bar r \left(\d \bar r\over \d t\right)_-  = 
8\pi \bar r \left(  H(t)\, \bar r  -  1 \right).
\end{equation}
Note that  $\dot S_-(t)$ changes sign at $\bar r(t) = H(t)^{-1}$. 
That is, an apparent horizon is present at the Hubble sphere  $\bar r_\mathrm{Hubble}(t) = H(t)^{-1}$ (sometimes called the ``speed of light sphere'').

One could also work in comoving coordinates where
\begin{equation}
S(t) = a(t)^2 \,4\pi \,r(t)^2,
\end{equation}
and 
\begin{equation}
\dot S_\pm(t) = 8\pi \, \{ a(t) \dot a(t) r(t)^2 + a(t)^2  r(t) \dot r_\pm(t)\} =  S(t)  \left\{ H \pm {1\over a(t) r(t)} \right\}. 
\end{equation}
There is again an apparent horizon when $\dot S_-=0$, at the same physical location where
\begin{equation}
 \bar r_\mathrm{Hubble}(t)= a(t) \, r_\mathrm{Hubble}(t) = H(t)^{-1}.
\end{equation}
This apparent horizon is emphatically not a causal horizon; there is no obstruction to crossing an apparent horizon.\footnote{Despite claims sometimes made in the literature, the Hubble radius is \emph{not} ``the distance light travels since the Big Bang". }
If one wishes to work in SI units, reinstating the speed of light, then 
\begin{equation}
 \bar r_\mathrm{Hubble}(t)= a(t) \; r_\mathrm{Hubble}(t) = {c\over H(t)}.
\end{equation}

\subsection{Particle horizon (causal horizon)}
\enlargethispage{20pt}
In contrast the particle horizon is a causal horizon determined by how far an outward moving light ray could move from its source (or equivalently how far an incoming light ray could move towards its reception point). For definiteness let us assume the light ray is emitted at time $t=0$ at location $\bar r=0$, then one is interested in solving the differential equation
\begin{equation}
{\d \bar r\over \d t} =  H(t)\, \bar r  +  1.
\end{equation} 
Equivalently
\begin{equation}
{\d \bar r\over \d t} -  {\dot a \over a}\, \bar r = a\; {\d (\bar r/a)\over \d t}   =  1,
\end{equation} 
whence
\begin{equation}
 {\d (\bar r/a)}   =  {\d t\over a},
\end{equation} 
This has the obvious solution ($t_*$ being the time of the Big Bang when $a(t_*)=0$) 
\begin{equation}
{\bar r(t)\over a(t)}    = r(t) =  \int_{t_*}^t {\d t\over a(t)} = \eta(t).
\end{equation} 
Equivalently
\begin{equation}
\bar r_\mathrm{particle}(t)   =   a(t)\; r_\mathrm{particle}(t) 
= a(t) \; \eta(t).
\end{equation} 
This particle horizon is by construction a causal horizon.
Note that the particle horizon has a very simple representation in terms of the conformal time coordinate. 

\subsection{Summary}
The apparent horizon (Hubble sphere) and particle horizon are distinct concepts, and can occur at radically different locations:
\begin{equation}
 \bar r_\mathrm{Hubble}(t)= a(t) \, r_\mathrm{Hubble}(t) = {c\over H(t)},
\end{equation}
\emph{versus}
\begin{equation}
\bar r_\mathrm{particle}(t)   =   a(t) \,r_\mathrm{particle}(t) = a(t) \int_{t_*}^t {\d t\over a(t)} = a(t) \, \eta(t).
\end{equation} 
In particular, since the Hubble sphere is not a causal horizon, one should \emph{not} attempt to apply ``causality'' arguments to the Hubble sphere.
\footnote{The occurrence of this particular misinterpretation of causality is distressingly common.}

What is always true based on elementary dimensional analysis is that
\begin{equation}
\bar r_\mathrm{particle}(t)  = \bar r_\mathrm{Hubble}(t)
\times (\hbox{dimensionless number}). 
\end{equation}
But there is absolutely no reason for this dimensionless number to be of order unity. 
In fact in the presence of cosmological inflation, whether it be exponential inflation, $a(t)\sim\exp(H_\mathrm{inflation}\, t)$, or power law inflation, $a(t) \sim t^n$ with $n\in(0,1]$, the integral $\eta(t) = \int_{t_*}^t {\d t\over a(t)}$ formally diverges, pushing the Big Bang out to negative infinity in conformal time, $\eta_* \to -\infty$, while pushing the particle horizon out to positive infinity,  $\bar r_\mathrm{particle}\to +\infty$. Even if cosmological inflation switches on and off at some finite time, the particle horizon can be made arbitrarily large compared to the Hubble radius.

\section{De Sitter spacetime}
The de Sitter spacetime is most typically presented in static coordinates:
\begin{equation}
\label{E:dS-static}
\d s^2 = -\left(1 - H^2 \bar r^2\right) \d \bar t^2 +
{\d \bar r^2 \over 1 - H^2 \bar r^2} + \bar r^2\d\Omega^2. 
\end{equation}
For this line element the Einstein tensor is $G_{ab} =-(3 H^2) g_{ab}$, corresponding to pure cosmological constant. 
Using the coordinate transformation
\begin{equation}
\bar t = t + \int {H \bar r \over1-H^2 \bar r^2} \; \d \bar r 
= t + {\ln(1-H^2 \bar r^2)\over 2H},
\end{equation}
we can cast de Sitter spacetime into Painlev\'e--Gullstrand form
\begin{equation}
\label{E:dS-PG}
\d s^2 = - \d t^2 +
[\d \bar r- H \bar r  \, \d t]^2  + \bar r^2\d\Omega^2.
\end{equation}
Finally, to make it abundantly clear that de Sitter spacetime is just a special case of FLRW spacetime, consider the specific coordinate transformation $\bar r = r \;e^{Ht}$, so that $\d\bar r = e^{Ht} (\d r + H r \d t)$, and use this to recast the Painlev\'e--Gullstrand form of the de~Sitter spacetime in the comoving form:
\begin{equation}
\label{E:dS-comoving}
\d s^2 = - \d t^2 +
e^{2Ht} \left\{ \d  r^2 +  r^2\d\Omega^2\right\} .
\end{equation}
Let us now generalize this discussion, first to the Kottler (Schwarzschild-de Sitter) spacetime, (which already presents a few subtleties), and then to the more complex and subtle McVittie spacetime.

\section{Kottler (Schwarzschild--de Sitter) spacetime}
\subsection{Standard form of Kottler}
The Kottler (Schwarzschild--de Sitter) spacetime is most typically presented in static coordinates~\cite{OSCO}:
\begin{equation}
\label{E:Kottler-static}
\d s^2 = -\left(1-{2m\over \bar r} - H^2 \bar r^2\right) \d \bar t^2 +
{\d \bar r^2 \over 1-{2m\over \bar r} - H^2 \bar r^2} + \bar r^2\d\Omega^2. 
\end{equation}
For this line element the Einstein tensor is $G_{ab} = -(3 H^2) g_{ab}$, corresponding to pure cosmological constant, plus a central point mass. 
The fiducial observers are in this situation best taken to be integral curves of the timelike Killing vector, and so are described by the non-geodesic 4-velocity field 
\begin{equation}
V^a = {1\over \sqrt{1-2m/\bar r-H^2\bar r^2}} \left(1, 0,0,0\right);
\qquad V_a = \sqrt{1-2m/\bar r-H^2\bar r^2}\; (-1,0,0,0).
\end{equation}
Here the 4-acceleration for this set of fiducial observers is
\begin{equation}
A^a = V^b \nabla_b V^a = \left(0, \,{m\over \bar r^2}- H^2 \bar r \,,0,0\right).
\end{equation}

\subsection{Five variant forms of Kottler}
Under suitable  coordinate changes, we first present three alternative Painlev\'e--Gullstrand-like formulations of the Kottler spacetime:
\begin{itemize}
\item 
Using the coordinate transformation
\begin{equation}
\bar t = t + \int {H \bar r \over\sqrt{1-2m/\bar r} (1-2m/\bar r+H^2 \bar r^2 )} \d \bar r \,,
\end{equation}
we cast the metric into the form
\begin{equation}
\label{E:PGK-1a}
\d s^2 = -\left(1-{2m\over \bar r}\right) \d t^2 +
{[\d \bar r- H \bar r \,\sqrt{1-2m/\bar r} \, \d t]^2 \over 1-{2m\over \bar r}} + \bar r^2\d\Omega^2,
\end{equation}
which we can also write as
\begin{equation}
\label{E:PGK-1b}
\d s^2 = -\left(1-{2m\over \bar r}\right) \d t^2 +
\left[{\d \bar r\over \sqrt{1-2m/\bar r}}- H \bar r \, \d t\right]^2 + \bar r^2\d\Omega^2.
\end{equation}
This form of the metric neatly disentangles the local physics, (depending only on the point mass $m$), from the cosmological physics, (depending only on the Hubble parameter $H$.
Specifically, as $m\to0$ this becomes de Sitter space in Painlev\'e--Gullstrand form (\ref{E:dS-PG}), whereas if $H\to 0$ this becomes Schwarzschild in standard form.

The fiducial observers (4-orthogonal to the spatial slices, so $V^\flat\propto \d t$) are in this situation described by the non-geodesic 4-velocity field 
\begin{equation}
V^a = \left({1\over \sqrt{1-2m/\bar r}}, Hr,0,0\right);
\qquad V_a = \sqrt{1-2m/\bar r}\; (-1,0,0,0).
\end{equation}
Here the 4-acceleration is 
\begin{equation}
A^a = V^b \nabla_b V^a = 
\left(0, \,{m\over \bar r^2} \, ,0,0\right).
\end{equation}
\item
Using the coordinate transformation
\begin{equation}
\bar t = t + \int {\sqrt{2m/\bar r} \over\sqrt{1-H^2 \bar r^2} \;(1-2m/\bar r+H^2 \bar r^2) } \d \bar r \,,
\end{equation}
we have another \emph{partial} Painlev\'e--Gullstrand form
\begin{equation}
\label{E:PGK-2a}
\d s^2 = -\left(1-H^2 \bar r^2\right) \d t^2 +
{[\d \bar r- \sqrt{2m/\bar r} \,\sqrt{1-H^2 \bar r^2} \, \d t]^2 \over 1-H^2 \bar r^2} + \bar r^2\d\Omega^2,
\end{equation}
which we can also write as 
\begin{equation}
\label{E:PGK-2b}
\d s^2 = -\left(1-H^2 \bar r^2\right) \d t^2 +
\left[{\d \bar r\over \sqrt{1-H^2 \bar r^2}} - \sqrt{2m/\bar r} \, \, \d t\right]^2 + \bar r^2\d\Omega^2,
\end{equation}
As $m\to0$ this becomes de Sitter in static 
form (\ref{E:dS-static}), whereas if $H\to 0$ this becomes Schwarzschild in Painlev\'e--Gullstrand form.

The fiducial observers (4-orthogonal to the spatial slices, so $V^\flat\propto \d t$) are in this situation described by the non-geodesic 4-velocity field 
\begin{equation}
V^a = \left({1\over \sqrt{1- H^2 \bar r^2}},\, \sqrt{2m\over \bar r},0,0\right);\qquad V_a = \sqrt{1-H^2 \bar r^2} \; (-1,0,0,0).
\end{equation}
Here the 4-acceleration is 
\begin{equation}
A^a = V^b \nabla_b V^a = 
\left(0, \,H^2\bar r \, ,0,0\right).
\end{equation}

\item
Using the coordinate transformation
\begin{equation}
\bar t = t + \int { \sqrt{2m/\bar r+H^2 \bar r^2}\over{1-2m/\bar r-H^2 \bar r^2}  } \;  \d \bar r\,,
\end{equation}
we have the \emph{full} Painlev\'e--Gullstrand form
\begin{equation}
\label{E:PGK-3a}
\d s^2 = - \d t^2 +
\left[\d \bar r- \sqrt{2m/\bar r + H^2 \bar r^2} \; \d t\right]^2 + \bar r^2\d\Omega^2.
\end{equation}
As $m\to0$ this becomes de Sitter in Painlev\'e--Gullstrand
form (\ref{E:dS-PG}), whereas if $H\to 0$ this becomes Schwarzschild in Painlev\'e--Gullstrand form.
The fiducial observers (4-orthogonal to the spatial slices, so $V^\flat\propto \d t$) are in this situation described by the geodesic 4-velocity field 
\begin{equation}
V^a = \left(1,\sqrt{2m/\bar r + H^2 \bar r^2},0,0\right);
\qquad V_a = (-1,0,0,0).
\end{equation}
It is easy to check that the 4-acceleration is zero: $A^a = V^b \nabla_b V^a =0$.

\end{itemize}
These four line elements  (\ref{E:Kottler-static})-(\ref{E:PGK-1b})-(\ref{E:PGK-2b})-(\ref{E:PGK-3a}) are all equally valid slicings of the Kottler spacetime; in all cases the Einstein tensor is $G_{ab} = -(3 H^2) g_{ab}$, corresponding to pure cosmological constant (in the presence of a central point mass). Depending on one's choice of slicing, one could make different choices of fiducial observer, focussing on different aspects of the physics. 

\clearpage
Finally to make it abundantly clear that Kottler spacetime is just a special case of Schwarzschild embedded in a specific FLRW (de Sitter) spacetime, consider the coordinate transformation $\bar r = r e^{Ht}$, so that $\d\bar r = e^{Ht} (\d r + H r \d t)$, and use this to recast the Painlev\'e--Gullstrand form of the Kottler spacetime (\ref{E:PGK-3a}) into the not entirely obvious comoving form:
\begin{equation}
\label{E:Kottler-comoving}
\d s^2 = - \d t^2 +
e^{2Ht} \left\{ \left[\d  r + \left(H r- \sqrt{2m e^{-3Ht} /r + H^2  r^2}\right) \, \d t\right]^2 +  r^2\d\Omega^2\right\} .
\end{equation}
Expanding, we have
\begin{eqnarray}
\label{E:Kottler-comoving1bbb}
\d s^2 &=& - \left\{1-  e^{2Ht} \left(H r- \sqrt{2m e^{-3Ht} /r + H^2  r^2}\right)^2 \right\} \d t^2
\nonumber\\
&&- 2 \left(H r- \sqrt{2m e^{-3Ht} /r + H^2  r^2}\right) \d r \d t
+ e^{2Ht} \left\{ \d  r^2 +  r^2\d\Omega^2\right\} .
\end{eqnarray}
It is relatively easy to explicitly check that the Einstein tensor is still $G_{ab} = - 3 H^2 g_{ab}$. 

In this form the connection between Kottler spacetime and  spatially flat $k=0$ FLRW is manifest since the limit $m\to0$ simply yields
\begin{equation}
\d s^2 = - \d t^2 +
e^{2Ht} \left\{ \d  r^2 +  r^2\d\Omega^2\right\} .
\end{equation}

The fiducial observers for (\ref{E:Kottler-comoving}) or (\ref{E:Kottler-comoving1bbb}) are 
described by the geodesic 4-velocity field 
\begin{equation}
V^a = \left(1, -\left[H r- \sqrt{2m e^{-3Ht} /r + H^2  r^2}\right], 0,0\right);\qquad V_a = (-1,0,0,0).
\end{equation}
It is relatively easy to check that the 4-acceleration is zero: $A^a = V^b \nabla_b V^a =0$.

In the same manner we can convert the (\ref{E:PGK-1b}) form of the Kottler spacetime  into a distinct  not entirely obvious comoving form
\begin{equation}
\label{E:Kottler-comoving1b}
\d s^2 = -\left(1-{2m e^{-Ht} \over r}\right) \d \bar t^2 +
e^{2Ht} \left\{ 
{\left(\d r+ Hr \left[1-\sqrt{1-2me^{-Ht} /r}\right]\d t\right)^2 \over 1-{2m e^{-Ht}\over r} } +  r^2\d\Omega^2\right\}. 
\end{equation}
It is relatively easy to explicitly check that the Einstein tensor is still $G_{ab} = - 3 H^2 g_{ab}$. Furthermore as $m\to0$ one recovers (\ref{E:dS-comoving}) the comoving slicing of the de~Sitter spacetime. 

\clearpage
The fiducial observers 4-orthogonal to the spatial slices are in this situation described by the non-geodesic 4-velocity field 
\begin{equation}
V^a = {\left(1,\, Hr \left[1-\sqrt{1-2me^{-Ht} /r}\right],0,0 \right)\over\sqrt{1- 2me^{-Ht} /r} };\qquad 
V_a = \sqrt{1- 2me^{-Ht} /r}\; (-1,0,0,0).
\end{equation}
Here the 4-acceleration is 
\begin{equation}
A^a = V^b \nabla_b V^a = 
\left(0, \,{m e^{-3Ht}\over \bar r^2} \, ,0,0\right).
\end{equation}

\subsection[Summary\\ \bigskip]{Summary}
Whereas the Kottler (Schwarzschild--de Sitter) spacetime is most commonly presented in static coordinates (\ref{E:Kottler-static}), it can with a little work be converted into Painlev\'e--Gullstrand form (\ref{E:PGK-1b})-(\ref{E:PGK-2b})-(\ref{E:PGK-3a}), and thence into comoving coordinates --- as per  (\ref{E:Kottler-comoving})--(\ref{E:Kottler-comoving1bbb})  and (\ref{E:Kottler-comoving1b}) above.  While finding the required coordinate transformations is relatively  straightforward, the process is not entirely obvious.

\section{McVittie spacetime}
The McVittie spacetime~\cite{McVittie, Kaloper:2010, Lake:2011, Faraoni:2012} is a perfect fluid spacetime that is as close as one can get to modelling  a Schwarzschild black hole embedded in an arbitrary FLRW spacetime.  

\subsection{Traditional form of McVittie spactime}
It is traditional to work in isotropic coordinates, where for $k=0$ the McVittie line element is given by the equivalent of ~\cite{McVittie}: 
\begin{equation}
\label{E:McVittie-standard}
\d s^2 = - \left(1-{m\over 2 a(t) \tilde r} \over 1+{m\over 2 a(t) \tilde r} \right)^2 \d t^2 
+ \left(1+{m\over 2 a(t) \tilde r} \right)^4 a(t)^2 
\{ \d \tilde r^2 + \tilde r^2 \d\Omega^2\}.
\end{equation}
\begin{itemize}
\item For $a(t)=1$ this is Schwarzschild spacetime in isotropic coordinates.
\item For $m=0$ this is a generic spatially flat $k=0$ FLRW spacetime.
\item{} While not entirely obvious, for $a(t) = e^{Ht}$ this is indeed Kottler (Schwarzschild-de~Sitter) spacetime in disguise. 
\end{itemize}
In these coordinates the fiducial observer (4-orthogonal to the spatial slices) has 4-velocity
\begin{equation}
V^a = \left(1+{m\over 2 a(t) \tilde r} \over 1-{m\over 2 a(t) \tilde r} \right) (1,0,0,0); 
\end{equation}
and the unit radial vector is
\begin{equation}
R^a = \left(1+{m\over 2 a(t) \tilde r} \right)^{-2} 
{1\over a(t)} \;(0, 1,0,0).
\end{equation}

Straightforward computation yields the orthonormal stress-energy components. The density is particularly simple,
\begin{equation}
\rho = {3\over8\pi} {\dot a^2\over a^2}=  {3\over8\pi} H^2,
\end{equation}
whereas the pressure is slightly more complicated
\begin{eqnarray}
p &=& {1\over8\pi} \left\{-
{2a\ddot a + \dot a^2\over a^2} 
-
{4m/[2a\tilde r]\over 1-m/[2a\tilde r]} \; {a\ddot a - \dot a^2\over a^2} 
\right\}
\nonumber\\
&=&
{1\over8\pi} \left\{ -3 H^2 - 2 \dot H 
-
{4m/[2a\tilde r]\over 1-{m\over 2a\tilde r}} \; {\dot H} 
\right\}
\nonumber\\
&=&
{1\over8\pi} \left\{ -3 H^2 
- 2 \; {1+{m\over 2a\tilde r}\over 1-{m\over 2a\tilde r}} \; {\dot H} 
\right\}.
\end{eqnarray}
All other components of the stress-energy are zero. 
Note that the energy density is identically that of FLRW, while the pressure asymptotes to that of FLRW. In view of the fact that there is a non-zero pressure gradient the fiducial observers, being in this situation defined by the fluid flow, will now not be geodesic. In fact the fiducial observers have  4-acceleration
\begin{equation}
A^a = V^b \nabla_b V^a =  {m\over a^2 \bar r^2 (1+{m\over 2a\tilde r})^3 (1-{m\over2a\tilde r})} \; R^a,
\end{equation}
and satisfy the Euler equation of fluid equilibrium
\begin{equation}
(\rho+p) A^a = - \left( g^{ab} + V^a V^b\right) \nabla_b p.
\end{equation}

\subsection{McVittie spacetime in Schwarzschild radial coordinates}

Kaloper--Kleban--Martin~\cite{Kaloper:2010} rewrite the McVittie line element by defining the Schwarzschild radial coordinate $\bar r$ by 
\begin{equation}
\bar r = \left(1+{m\over 2 a(t) \tilde r} \right)^2 a(t) \tilde r,
\end{equation}
and transforming the line element into the equivalent of 
\begin{equation}
\label{E:McVittie-KKM-a}
\d s^2 =
- \left(1-{2m\over \bar r}\right) \d t^2 
+ {[\d \bar r - \sqrt{1-2m/\bar r} \; H(t) \bar r \;\d t]^2\over 1-2m/\bar r}
+ \bar r^2 \d\Omega^2.
\end{equation}
Let us rewrite this as
\begin{equation}
\label{E:McVittie-KKM-b}
\d s^2 =
- \left(1-{2m\over \bar r}\right)  \d t^2 
+ \left[{\d \bar r\over\sqrt{1-2m/\bar r}} -  H(t) \bar r \d t\right]^2
+ \bar r^2 \d\Omega^2.
\end{equation}
This form of the metric again neatly disentangles the local physics, (depending only on the point mass $m$), from the cosmological physics, (depending only on the Hubble parameter $H(t)$, which is now allowed to be time-dependent).
Specifically,  setting $H(t)\to H$ yields equation (\ref{E:PGK-1b}), one of the representations of Kottler spacetime, while setting $m\to0$ yields equation (\ref{E:FLRW5}) one of the Painlev\'e--Gullstrand representations of $k=0$ FLRW spacetime.

The Eulerian observer has 4-velocity
\begin{equation}
V_a = \sqrt{1-2m/r} \; (-1,0,0,0); \qquad V^a = \left({1\over\sqrt{1-2m/r}},H(t) r,0,0\right);
\end{equation}
and the unit radial vector is
\begin{equation}
R^a = \sqrt{1-2m/r}(0, 1,0,0); \qquad 
R_a = \left(- H(t) r,{1\over\sqrt{1-2m/r}},0,0 \right).
\end{equation}
The density and pressure are now
\begin{equation}
\rho={3H(t)^2\over8\pi}; \qquad
p = -\rho + {\dot H(t)\over 4\pi\sqrt{1-2m/r}}.
\end{equation}
The fiducial observers have non-zero 4-acceleration
\begin{equation}
A^a = \left( 0, \, {m\over r^2} \,, 0,0  \right),
\end{equation}
and satisfy the Euler equation of fluid equilibrium
\begin{equation}
(\rho+p) A^a = - \left( g^{ab} + V^a V^b\right) \nabla_b p.
\end{equation}

\subsection{McVittie spacetime in comoving radial coordinates}

Now set $\bar r = a(t) r$, so that $\d\bar r = a(t)\,  (\d r + H(t) r  \d t) $. Then
\begin{equation}
\d s^2 =
- \left(1-{2m\over a(t) r}\right) \d t^2 
+ \left[{a(t)  (\d r + H(t) r  \d t)\over\sqrt{1-{2m\over a(t) r}}} -  H(t) a(t) r \d t\right]^2
+ a(t)^2 r^2 \d\Omega^2.
\end{equation}
Thence we obtain a comoving form of the McVittie spacetime
\begin{equation}
\label{E:McVittie-comoving}
\d s^2 =
- \left(1-{2m\over a(t) r}\right) \d t^2 
+ a(t)^2 \left\{ \left[{\left(\d r + H(t) r \left[1-\sqrt{1-{2m\over a(t) r}}\right]  \d t\right)^2\over{1-{2m\over a(t) r}}} \right]
+  r^2 \d\Omega^2 \right\}.
\end{equation}
\begin{itemize}
\item For $a(t)=1$, so that $H(t)=0$,  this is Schwarzschild spacetime in standard coordinates.
\item For $a=e^{Ht}$, so that $H(t)\to H$,  this is equivalent to the (\ref{E:Kottler-comoving1b}) representation of Kottler spacetime. 
\item For $m=0$ this is standard $k=0$ FLRW spacetime in comoving coordinates.
\end{itemize}

The natural Eulerian observer (the closest you can get to defining the Hubble flow) is specified by the unit 4-vector
\begin{equation}
V^a = {\left(1, -H(t) r \left[1-\sqrt{1-{2m\over a(t) r}}\,\right]  ,0,0\right)\over 1 - {2 m\over a(t) r} }. 
\end{equation}
This corresponds to the covector
\begin{equation}
V_a = \sqrt{1-{2m\over a(t) r}} \; (-1,0,0,0).
\end{equation}

The unit radial 4-vector is
\begin{equation}
R^a = {\sqrt{1-{2m\over a(t) r}} \over a(t)} \;\;(0,1,0,0).
\end{equation}
This corresponds to the covector
\begin{equation}
R_a =  {a(t)\over  \sqrt{1-{2m\over a(t) r}} }  \left(H(t) r \left[1-\sqrt{1-{2m\over a(t) r}}\,\right]  ,1 ,0,0 \right) .
\end{equation}
In the appropriate orthonormal basis the energy density is still
\begin{equation}
\rho = {3\over8\pi} {\dot a^2\over a^2}=  {3\over8\pi} H(t)^2
\end{equation}
while the pressure now becomes
\begin{eqnarray}
p &=& {1\over8\pi} \left\{ -3 H(t)^2  +  2 \; {\dot H(t)\over \sqrt{1 - {2 m\over a(t) r} }} 
\right\}
\end{eqnarray}
All other components of the stress-energy are zero. 
The fiducial observers have non-zero 4-acceleration
\begin{equation}
A^a = \left( 0, \, {m\over r^2 a(t)^3} \,, 0,0  \right),
\end{equation}
and satisfy the Euler equation of fluid equilibrium
\begin{equation}
(\rho+p) A^a = - \left( g^{ab} + V^a V^b\right) \nabla_b p.
\end{equation}

\subsection{McVittie spacetime in (conformal) Painlev\'e--Gullstrand  form}

On quite general grounds, (since McVittie spacetime is spherically symmetric, does not possess any wormhole throats, and has a non-negative Misner--Sharp quasi-local mass),   a (full) Painlev\'e--Gullstrand  form for McVittie spacetime must exist~\cite{Nielsen:2005}. But as Faraoni has pointed out~\cite{Faraoni:2012, Faraoni:2020},  that (full) Painlev\'e--Gullstrand  form  depends on a quite messy (and implicit) integrating factor, over which one has little to no control; making the  (full) Painlev\'e--Gullstrand  form completely explicit seems a formidable task.
Fortunately, there is an intermediate step, a \emph{conformal} Painlev\'e--Gullstrand  form, that is much easier to make fully explicit. 

Start with McVittie spacetime in traditional form:
\begin{equation}
\d s^2 = - \left(1-{m\over 2 a(t) \tilde r} \over 
1+{m\over 2 a(t) \tilde  r} \right)^2 \d t^2 
+ \left(1+{m\over 2 a(t) \tilde r} \right)^4 a(t)^2 
\{ \d \tilde r^2 + \tilde r^2 \d\Omega^2\}.
\end{equation}
Define $\bar r = a(t) \tilde  r$. Then as usual $\d \tilde  r = \d(\bar r /a) = (\d\bar r - H(t)\, \bar r\, \d t)/a$, and so
\begin{equation}
\d s^2 = - \left(1-{m\over 2 \bar r} \over 1+{m\over 2 \bar r} \right)^2 \d t^2 
+ \left(1+{m\over 2 \bar r} \right)^4  
\{ [\d \bar r- H(t)\,\bar r\,\d t]^2 + \bar r^2 \d\Omega^2\}.
\end{equation}

\clearpage
This is not quite of Painlev\'e--Gullstrand  form; but it is \emph{conformal} to Painlev\'e--Gullstrand  form:
\begin{equation}
\label{E:McVittie-CPG}
\d s^2 =   \left(1+{m\over 2 \bar r} \right)^4   \left\{ - \left([1-{m\over 2 \bar r}]^2 \over [1+{m\over 2 \bar r}]^6 \right) \d t^2 
+
\{ [\d \bar r- H(t)\,\bar r\,\d t]^2 + \bar r^2 \d\Omega^2\}\right\} .
\end{equation}
Note the spatial slices are conformally flat, and both the conformal factor and lapse function are time independent --- the only time-dependence has now been isolated in the Hubble parameter $H(t)$. 
Straightforward computation yields the 
temporal and radial legs of the tetrad
\begin{equation}
V^a = {1 +{m\over2r}\over1 -{m\over2r}}\; 
\Big(1,\,H(t) r,0,0\Big); \qquad 
R^a = \left(0,\, {1\over 1 +{m\over2r}},0,0 \right).
\end{equation}
The orthonormal stress-energy components are:
\begin{equation}
\rho = {3\over8\pi} H(t)^2;
\qquad
p = {1\over8\pi} \left\{-
3 H(t)^2 - 2
{1+{2m\over\bar r}\over 1-{m\over2\bar r}} \; \dot H(t)
\right\}.
\end{equation}
All other components of the stress-energy are zero. 
As required this is a perfect fluid. As required, as $H(t)\to H$, so that $\dot H=0$,  one recovers the isotropic form of Kottler spacetime.

\subsection{Summary}

We have now extracted 4 equivalent forms of the McVittie spacetime --- the traditional (\ref{E:McVittie-standard}), the Schwarzschild variant (\ref{E:McVittie-KKM-a})--(\ref{E:McVittie-KKM-b}), a comoving variant (\ref{E:McVittie-comoving}), and finally a conformally Painlev\'e--Gullstrand (\ref{E:McVittie-CPG}) variant.

\section{Discussion}

\enlargethispage{20pt}
Overall, we have seen that coordinate freedom in cosmology can be used to repackage and reorganize standard cosmological models in multiple different ways. This repackaging and reorganization can often simplify \emph{some} aspects of the physics, while making \emph{other} aspects seem more (apparently) complex. 
\begin{itemize}
\item We have explored three specific ways of rewriting  the generic $k=0$ FLRW cosmologies; equations (\ref{E:FLRW1}), (\ref{E:FLRW3}), (\ref{E:FLRW5}), and their Cartesian versions
(\ref{E:FLRW2}), (\ref{E:FLRW2}), (\ref{E:FLRW6}), there are many others. The three we have explored either make the Hubble flow simple, or make the light cones simple, or make the spatial slices simple. But there is no free lunch; the underlying physics is invariant. 

\item 
We have similarly considered three versions of de~Sitter space, (\ref{E:dS-static}), (\ref{E:dS-PG}), (\ref{E:dS-comoving}); either making the spacetime manifestly static, or making the spatial slices flat, or making the connection to generic FLRW manifest. 

\item 
For the Kottler spacetime we have developed six different line elements, (\ref{E:Kottler-static}), (\ref{E:PGK-1a}), (\ref{E:PGK-2a}), (\ref{E:PGK-3a}), (\ref{E:Kottler-comoving}), (\ref{E:Kottler-comoving1b}); focussing on different aspects of the physics. One either makes the spacetime manifestly static, or has three ways to make the spatial slices relatively simple, or has two ways to make the connection to generic FLRW manifest. There are yet other possibilities that one might explore.

\item 
For the McVittie spacetime we presented four different line elements, (\ref{E:McVittie-standard}),  (\ref{E:McVittie-KKM-a}),  (\ref{E:McVittie-comoving}),  (\ref{E:McVittie-CPG}), two of which seem to be novel. 
The traditional version (\ref{E:McVittie-standard}) is spatially isotropic, but every nonzero metric component is explicitly time dependent. The Schwarzschild version (\ref{E:McVittie-KKM-a}) sets three metric components to be time independent, and tightly constrains the time dependence of the  remaining terms. The ``comoving'' line element (\ref{E:McVittie-comoving}) makes the connection with generic $k=0$ FLRW manifest, while the \emph{conformally} Painlev\'e--Gullstrand version (\ref{E:McVittie-CPG}) makes the spatial slices time independent and eliminates explicit occurrences of the scale factor $a(t)$ in favour of the Hubble parameter $H(t)$. 
\end{itemize}

Perhaps the most bizarre feature of the above discussion is that one can apparently eliminate the expansion of the universe with a suitable choice of coordinates; of course there is then a \emph{different} price to play --- the light cones then ``tip over'' and one must be \emph{much} more careful when deciding which trajectories are now to be regarded as ``superluminal''. Specifically, the variant presentations of the Kottler and McVittie line elements give one a much better handle on how to how to merge the gravitational field of a non-perturbative localized compact object with the Hubble flow of an asymptotically FLRW cosmology. 
Finally we point out that familiarity with these variant coordinate systems is also helpful  in understanding the symmetries (both explicit and hidden); and  in demystifying the horizon structure. 
Ultimately we would be interested in extending these ideas to generic non-perturbative deviations from FLRW cosmology.

\bigskip
\bigskip
\hrule\hrule\hrule

\clearpage
\appendix
\section{Appendix: Collected formulae}

Here we present some collected formulae for easy reference.

\subsection{Spatially flat FLRW}

By suitable choice of coordinates spatially flat FLRW spacetime can be represented in any of the following six equivalent forms:
\begin{equation}
\label{E:FLRW1A}
\d s^2 = - \d t^2 + a(t)^2 \{ \d r^2 + r^2 \d\Omega^2\},
\end{equation}
\begin{equation}
\label{E:FLRW2A}
\d s^2 = - \d t^2 + a(t)^2 \{ \d x^2 + \d y^2 + \d z^2\}.
\end{equation}

\begin{equation}
\label{E:FLRW3A}
\d s^2 =  a(\eta)^2 \{ -\d \eta^2 + \d r^2 + r^2 \d\Omega^2\}.
\end{equation}
\begin{equation}
\label{E:FLRW4A}
\d s^2 = a(\eta)^2 \{ - \d \eta^2 +  \d x^2 + \d y^2 + \d z^2\}.
\end{equation}

\begin{equation}
\label{E:FLRW5A}
\d s^2 = - \d t^2 + \{ [\d \bar r - H(t)\, \bar r \,\d t]^2 + \bar r^2 \d\Omega^2\}.
\end{equation}
\begin{equation}
\label{E:FLRW7A}
\d s^2 = -  \d t^2 
 + \left\{ [\d \bar x- H(t) \,\bar x\, \d t]^2 
 + [\d \bar y - H(t) \,\bar y \,\d t]^2 
 + [\d\bar z - H(t) \,\bar z \, \d t]^2\right\}.
\end{equation}

\subsection{Kottler}

By suitable choice of coordinates Kottler (Schwarzschild-de Sitter) spacetime  can be represented in any of the following six equivalent forms:
\begin{equation}
\label{E:Kottler-staticA}
\d s^2 = -\left(1-{2m\over \bar r} - H^2 \bar r^2\right) \d \bar t^2 +
{\d \bar r^2 \over 1-{2m\over \bar r} - H^2 \bar r^2} + \bar r^2\d\Omega^2. 
\end{equation}

\begin{equation}
\label{E:PGK-1aA}
\d s^2 = -\left(1-{2m\over \bar r}\right) \d t^2 +
{[\d \bar r- H \bar r \,\sqrt{1-2m/\bar r} \, \d t]^2 \over 1-{2m\over \bar r}} + \bar r^2\d\Omega^2,
\end{equation}
\begin{equation}
\label{E:PGK-2aA}
\d s^2 = -\left(1-H^2 \bar r^2\right) \d t^2 +
{[\d \bar r- \sqrt{2m/\bar r} \,\sqrt{1-H^2 \bar r^2} \, \d t]^2 \over 1-H^2 \bar r^2} + \bar r^2\d\Omega^2,
\end{equation}
\begin{equation}
\label{E:PGK-3aA}
\d s^2 = - \d t^2 +
\left[\d \bar r- \sqrt{2m/\bar r + H^2 \bar r^2} \; \d t\right]^2 + \bar r^2\d\Omega^2.
\end{equation}

\begin{equation}
\label{E:Kottler-comovingA}
\d s^2 = - \d t^2 +
e^{2Ht} \left\{ \left[\d  r + \left(H r- \sqrt{2m e^{-3Ht} /r + H^2  r^2}\right) \, \d t\right]^2 +  r^2\d\Omega^2\right\} .
\end{equation}
\begin{equation}
\label{E:Kottler-comoving1bA}
\d s^2 = -\left(1-{2m e^{-Ht} \over r}\right) \d \bar t^2 +
e^{2Ht} \left\{ 
{\left(\d r+ Hr \left[1-\sqrt{1-2me^{-Ht} /r}\right]\d t\right)^2 \over 1-{2m e^{-Ht}\over r} } +  r^2\d\Omega^2\right\}. 
\end{equation}

\subsection{McVittie}

By suitable choice of coordinates McVittie spacetime  can be represented in any of the following four equivalent forms:
\begin{equation}
\label{E:McVittie-standardA}
\d s^2 = - \left(1-{m\over 2 a(t) \tilde r} \over 1+{m\over 2 a(t) \tilde r} \right)^2 \d t^2 
+ \left(1+{m\over 2 a(t) \tilde r} \right)^4 a(t)^2 
\{ \d \tilde r^2 + \tilde r^2 \d\Omega^2\}.
\end{equation}

\begin{equation}
\label{E:McVittie-KKM-bA}
\d s^2 =
- \left(1-{2m\over \bar r}\right)  \d t^2 
+ \left[{\d \bar r\over\sqrt{1-2m/\bar r}} -  H(t) \bar r \d t\right]^2
+ \bar r^2 \d\Omega^2.
\end{equation}

\begin{equation}
\label{E:McVittie-comovingA}
\d s^2 =
- \left(1-{2m\over a(t) r}\right) \d t^2 
+ a(t)^2 \left\{ \left[{\left(\d r + H(t) r \left[1-\sqrt{1-{2m\over a(t) r}}\right]  \d t\right)^2\over{1-{2m\over a(t) r}}} \right]
+  r^2 \d\Omega^2 \right\}.
\end{equation}

\begin{equation}
\label{E:McVittie-CPGA}
\d s^2 =   \left(1+{m\over 2 \bar r} \right)^4   \left\{ - \left([1-{m\over 2 \bar r}]^2 \over [1+{m\over 2 \bar r}]^6 \right) \d t^2 
+
\{ [\d \bar r- H(t)\bar r\d t]^2 + \bar r^2 \d\Omega^2\}\right\} .
\end{equation}

\bigskip
\hrule\hrule\hrule
\section*{Acknowledgements}

RG was supported by a Victoria University of Wellington PhD Doctoral Scholarship
and was also indirectly supported by the Marsden Fund, 
via a grant administered by the Royal Society of New Zealand.
\\
MV was directly supported by the Marsden Fund, 
via a grant administered by the Royal Society of New Zealand.

\bigskip
\hrule\hrule\hrule

\clearpage

\end{document}